\documentclass[conference]{IEEEtran}

\usepackage{subfig}
\usepackage{slashbox}
\usepackage{array}
\usepackage{graphicx}
\usepackage{amsmath}
\usepackage{algorithmic}
\usepackage{algorithm}
\usepackage{doi}
\usepackage{url}
\usepackage{setspace}
\usepackage{enumitem}
\usepackage{tabularx}
\usepackage{graphicx,lipsum}
\usepackage{caption}
\usepackage{verbatim}

\newcolumntype{P}[1]{>{\centering\arraybackslash}p{#1}} 
\newcolumntype{M}[1]{>{\centering\arraybackslash}m{#1}} 
\begin{document}
\bibliographystyle{IEEEtran}
%

\title{Near Optimal Channel Assignment for Interference Mitigation in Wireless Mesh Networks}

\author{\IEEEauthorblockN{Ranadheer Musham, Srikant Manas Kala, Pavithra Muthyap, Pavan Kumar Reddy Mule,  and Bheemarjuna Reddy Tamma\textsuperscript \dag}
\IEEEauthorblockA{ Indian Institute of Technology Hyderabad, India\\
Email: [cs12b1026, cs12m1012, cs11b025, cs12b1025,  tbr\textsuperscript \dag]@iith.ac.in}}

\maketitle

\begin{abstract}
In multi-radio multi-channel (MRMC) WMNs, interference alleviation is affected through several network design techniques 
e.g., channel assignment (CA), link scheduling, routing  etc., intelligent CA schemes being the most effective tool for interference mitigation. 
CA in WMNs is an NP-Hard problem, and makes optimality a desired yet elusive goal in real-time deployments which are characterized by fast transmission and switching times and minimal end-to-end latency. 
The trade-off between optimal performance and minimal response times is often achieved through CA schemes that employ heuristics to propose efficient solutions. 
WMN configuration and physical layout are also crucial factors which decide network performance, and it has been demonstrated in numerous research works
that rectangular/square grid WMNs outperform random or unplanned WMN deployments in terms of network capacity, latency, and network resilience. 
In this work, we propose a smart heuristic approach to devise a near-optimal CA algorithm for grid WMNs (NOCAG). 
We demonstrate the efficacy of NOCAG by evaluating its performance against the minimal-interference CA generated through a rudimentary brute-force technique (BFCA), for the same WMN configuration. 
We assess its ability to mitigate interference both, theoretically (through interference estimation metrics) and experimentally (by running rigorous simulations in NS-3). 
We demonstrate that the performance of NOCAG is almost
as good as the BFCA, at a minimal computational overhead of O(n) compared to the exponential of BFCA.
\end{abstract}

\section{Introduction}

Wireless Networks have been one of the most common modes of usage for Internet and intranet these days. 
The number of users using wireless technologies are increasing exponentially every year because of the benefits of low-cost availability, increased mobility, and scalability.

Wireless Mesh Networks (WMNs) form the backbone of the next generation communication in areas with high population density, and corporate societies
because of their ease of integration with modern technologies viz., IEEE 802.11 Wireless local area networks (WLANs), LTE/4G and 5G through a single platform \cite{12Capone}. 

Multi-Radio Multi-Channel (MRMC) WMNs are advanced technological forms of the WMNs i.e., nodes have multiple radios  and there are multiple orthogonal channels available  for communication.
The performance of these MRMC WMNs is decided by the factors like network topology, channel assignment (CA) and routing.
We will be focusing on the CA problem because a good CA will affect the performance most.

\begin{figure}
  \centering%
  {\includegraphics[width=0.6\linewidth]{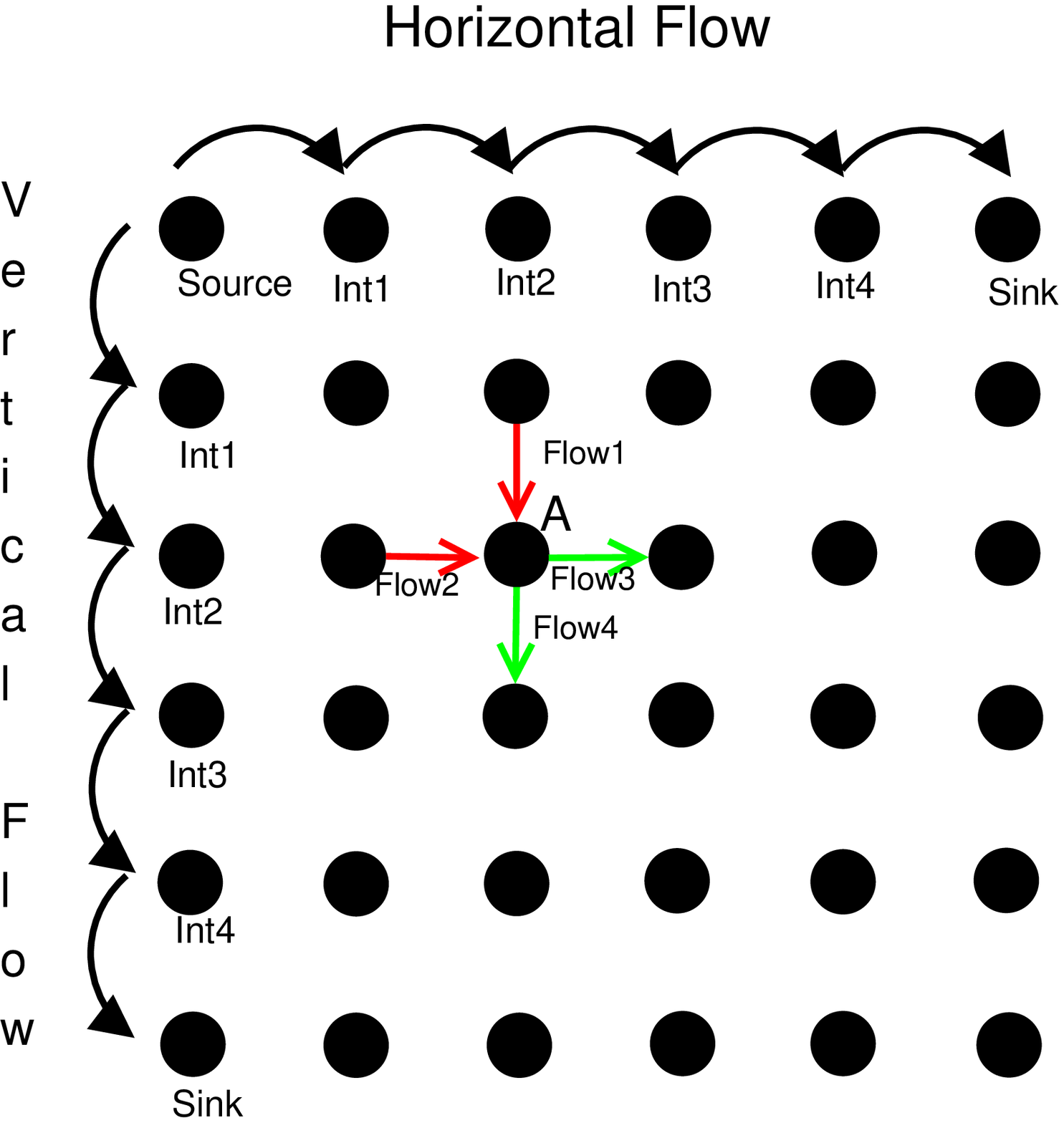}}
    \caption{Sample 6$\times$6 Grid WMN }
     \label{sample}
\end{figure} 

\section{Motivation and Related research work}
In our recent research, we have shown that radio co-location on a node leads to a special case of interference in WMNs 
and proposed  a novel approach to generate conflict graph \cite{kala2016interference}. 
Later on \cite{kala2015radio},  authors have come up with radio co-location aware CA algorithm  that works for any random WMNs 
and have achieved fairly good results than the existing algorithms. 

There are several heuristics proposed on the CA problem.  
Connected Low Interference Channel Assignment (CLICA) \cite{marina2010topology} uses a depth first search approach
to assign channels,
Topology-controlled Interference-aware Channel Assignment (TICA) \cite{chaudhry2012improving} is another approach that 
uses topology to assign channels by constructing Shortest Path Tree (SPT).
Genetic Algorithm (GA) \cite{sridhar2009channel} is a population-based stochastic search approach to assign channels to WMN links.

\subsection{Grid WMNs Vs. Random WMNs}

We chose the grid WMN because it has fairly good coverage area and network capacity as demonstrated in \cite{Grid,li2008gateway} and  
easy for the visualizing. In \cite{Grid}, it was shown that a grid WMN has almost double the network capacity than a random WMN. 
Grid WMNs are also better for gateway placement strategies to achieve high overall throughput \cite{li2008gateway,kala2015reliable}.
A sample grid WMN is shown in Figure~\ref{sample}. 

Centralized Channel Assignment (CCA) \cite{raniwala2004centralized} is a CA algorithm specially designed for grid WMNs. One major flaw with CCA is that it has no control over
choosing under-used channel so, there is a chance of over-utilizing
one particular channel leading to interference which is shown in following sections.

In this work we develop an intelligent and easily implementable heuristic algorithm to assign channels for grid WMNs, 
Near Optimal CA for Grids (NOCAG) which performs closer to the Brute Force computed CAs (BFCA) i.e., its performance is close to the optimum achievable.

\section{Proposed Work}

The network topology of a WMN can be represented as a graph.
Let $G_{WMN}=(V_{WMN},E_{WMN})$ represent MRMC WMN consisting of $m$ nodes, \
where $V_{WMN}$ denotes the set of nodes in the WMN and 
$E_{WMN}$ denotes the set of wireless links between nodes which lie within each other's transmission range. \
Each node has identical radios and number of radios on each need not be constant.

Let $CS$ be the set of available channels. \
and $CS_i$ represents the set of channels that are assigned to the radios on $i^{th}$ node.
$cs_{max}$ is the maximum  number of available orthogonal channels.
Let $R$ be the maximum number of radios on nodes i.e., $R_i$ represents the maximum number of radios on node $i$.
$|CS_i|$ denote the cardinality of the set $CS_i$ i.e., number of radios assigned a channel on node $i$.
The aim is to assign each node a subset of channels such that the WMN topology is preserved and \
interference is reduced so as to increase the WMN performance. The algorithm is presented in Algorithm~\ref{radcol}.

\renewcommand{\algorithmicrequire}{\textbf{Input:}}
\renewcommand{\algorithmicensure}{\textbf{Output:}}
\begin{algorithm}[htb!] 
\caption{Near Optimal Channel Assignment for Grid}
\label{radcol}
\begin{algorithmic}[1]
{\fontsize{9}{10}
\REQUIRE $G_{WMN} = (V_{WMN},E_{WMN})$, $R$,\\
      \hspace{.35cm}   $CS =\{Ch_{1}, Ch_{2},..., Ch_{cs_{max}}\}$
\ENSURE Channel Assignment NOCAG \\
\line(1,0){236}
\FOR {$i \in V_{WMN}$}
\FOR {$j \in Adj_i$}
\IF {$ CS_i \cap CS_j \ne \phi$}
\STATE $continue$
 \ENDIF
\IF	{$|CS_i|$ $<$ $R_i$ \&\& $|CS_j|$ $<$ $R_j$}
	\STATE $Ch$  $\leftarrow$ $CS$ - $CS_i$ - $CS_j$ 
	\STATE $Ch_{refine} $ $\leftarrow$ $Ch$ - $CS_{Adj_i}$
	\IF{$Ch_{refine} \ne \phi $}
	    \STATE $CS_{i}$ $\leftarrow$ $CS_{i}$ $\cup$ $k$;\\ $CS_{j}$ $\leftarrow$ $CS_{j}$ $\cup$ $k$; where $k$ $\in$ $Ch_{refine}$
	\ELSE 
	    \STATE $CS_{i}$ $\leftarrow$ $CS_{i}$ $\cup$ $k$; \\$CS_{j}$ $\leftarrow$ $CS_{j}$ $\cup$ $k$; where $k$ $\in$ $Ch$
	\ENDIF
	
 \ENDIF \\
 \COMMENT{Choose a channel that is not assigned to any of the radios on either node, then assign it to the free radio on each node.}
 \IF	{$|CS_i|$  $<$ $R_i$ \&\& $|CS_j|$  $=$ $R_j$}
	\STATE $Ch$ $\leftarrow$ {$CS -  CS_{Adj_i}$} $\cap$ $CS_j$ 
	\IF{$ Ch = \phi$}
	  \STATE $CS_{i}$ $\leftarrow$ $CS_{i}$ $\cup$ $k$ where $k$ $\in$ $Ch$ 
	\ELSE
	  \STATE $CS_{i}$ $\leftarrow$ $CS_{i}$ $\cup$ $k$ where $k$ $\in$ $CS_{j}$
	\ENDIF

 \ENDIF\\
 \COMMENT {Assign channel to the radio on first node from one of the channels that is assigned to second node.}
 \IF	{$|CS_i|$  $=$ $R_i$ \&\& $|CS_j|$ $<$ $R_j$}
	\STATE $Ch$ $\leftarrow$ {$CS -  CS_{Adj_j}$} $\cap$ $CS_i$ 
	\IF{$ Ch = \phi$}
	  \STATE $CS_{j}$ $\leftarrow$ $CS_{j}$ $\cup$ $k$ where $k$ $\in$ $Ch$ 
	\ELSE
	  \STATE $CS_{j}$ $\leftarrow$ $CS_{j}$ $\cup$ $k$  where  $k$ $\in$ $CS_{i}$
	\ENDIF
 \ENDIF\\
 \IF	{$|CS_i|$  $=$ $R_i$ \&\& $|CS_j|$  $=$ $R_j$}
	\STATE $k \in Ch_i$ $|$ $k$ $is$ $least$ $occured$ $in$ $CS_{Adj_j}$
	\STATE $l \in Ch_j$ $|$ $l$ $is$ $least$ $occured$ $in$ $CS_{Adj_i} $
	\STATE $CS_j \leftarrow CS_j - l + k$
 \ENDIF
\ENDFOR
\ENDFOR
}
\end{algorithmic}
\end{algorithm}

\subsection{Conceptual Background }

The algorithm is a novel approach in a sense that it does not use conflict graph, or an interference estimation metric like 
Total Interference Degree (TID), or $CXLS_{wt}$ \cite{kala2015reliable} 
to assign channels.  

We first discuss some of the aspects of interference prevalent in WMNs that need to be considered while assigning channels. 
From these considerations, we derive some crucial design components of our algorithm.

\subsubsection{Radios  on a node assigned same channel}                                      
It is not beneficial to assign the same channel to two or more radios on a single node \cite{kala2016interference},
as it becomes a source of interference and impacts the overall network performance. 
When a node transmits data to another node on a common channel and multiple radios on the  second node are also assigned the channel on which the first
node is transmitting, there is a high probability of radio co-location interference (RCI). RCI is detrimental to network performance \cite{kala2016interference}. 

\subsubsection{Connected nodes having pair of common channels}

This arises when a pair of connected nodes have more than one common channel to communicate. 
This also leads to a special case of interference observed in \cite{kala2016interference}, 
and this is a potential cause for throughput degradation.

\subsection{Step Wise Procedure}

The input given to the algorithm is the network topology, 
number of radios on  each node and the number of available orthogonal channels which are quite sufficient, 
 and necessary inputs for any channel assignment algorithm to assign channels in a WMN.
 
Algorithm takes a node and considers all the nodes adjacent to it. 
The behavior of the algorithm is based on the possible different scenarios on which the nodes considered are present on. 
The scenarios are enumerated below:

\begin{enumerate}
 \item \textbf{More than one common channel on the nodes:}
        This case does not arise as the algorithm make stepwise progress and at no point it will assign more than one common channel.
        For a pair of neighboring nodes.
  \item \textbf{Only one common channel on the nodes:}        
        There would be no change of channel assignment for any radio on either nodes.
   \item  \textbf{No common channel between the nodes and both nodes have at least one unassigned radio:}
        It assigns a common channel to one of the unassigned radio on each node such that conclusions drawn above are not violated.

   \item \textbf{No common channel between the nodes and only one of the nodes have unassigned radio:}
        Assigns a channel to the unassigned radio that is same as the channel assigned to radio on the other node. 
        If more than one channel is possible to assign then it is wise to choose the one which decreases the interference most.

   \item \textbf{No common channel between the nodes and neither have unassigned radio:}
         It tries to change the channel on one of the radios in such a way that both nodes can communicate and the rise in the interference is least.

\end{enumerate}
\subsection{Time Complexity Analysis} 
For a given grid WMN of $n \times n$ size, let $m$ be the total number of nodes i.e., $m$ = $n^{2}$, $k$ be the average number of radios on each node and
$c$ be the number of available channels.
Time complexity for computing a BFCA i.e., checking all the possible CAs and choosing the best CA is O($c^{(m*k)}$).

NOCAG chooses each node at a time and for each node it considers only its adjacent nodes.
Now for each node in a grid WMN maximum number of adjacent nodes can be 4, and in the worst case it checks for all available channels.
So the time complexity of the algorithm is O($4*m*c$) i.e., O($m*c$).
In general  $c$ is very low as compared to a normal WMN or $c < < m$. So the time complexity can depicted as O($m$).
The algorithm proposed is linear in terms of number of nodes in the WMN which is very good for any CA algorithm.

\subsection{Walk Through Example}

\begin{figure}
  \centering%
  \begin{tabular}{cc}
  \subfloat[Initial Grid]{\includegraphics[width=.4\linewidth]{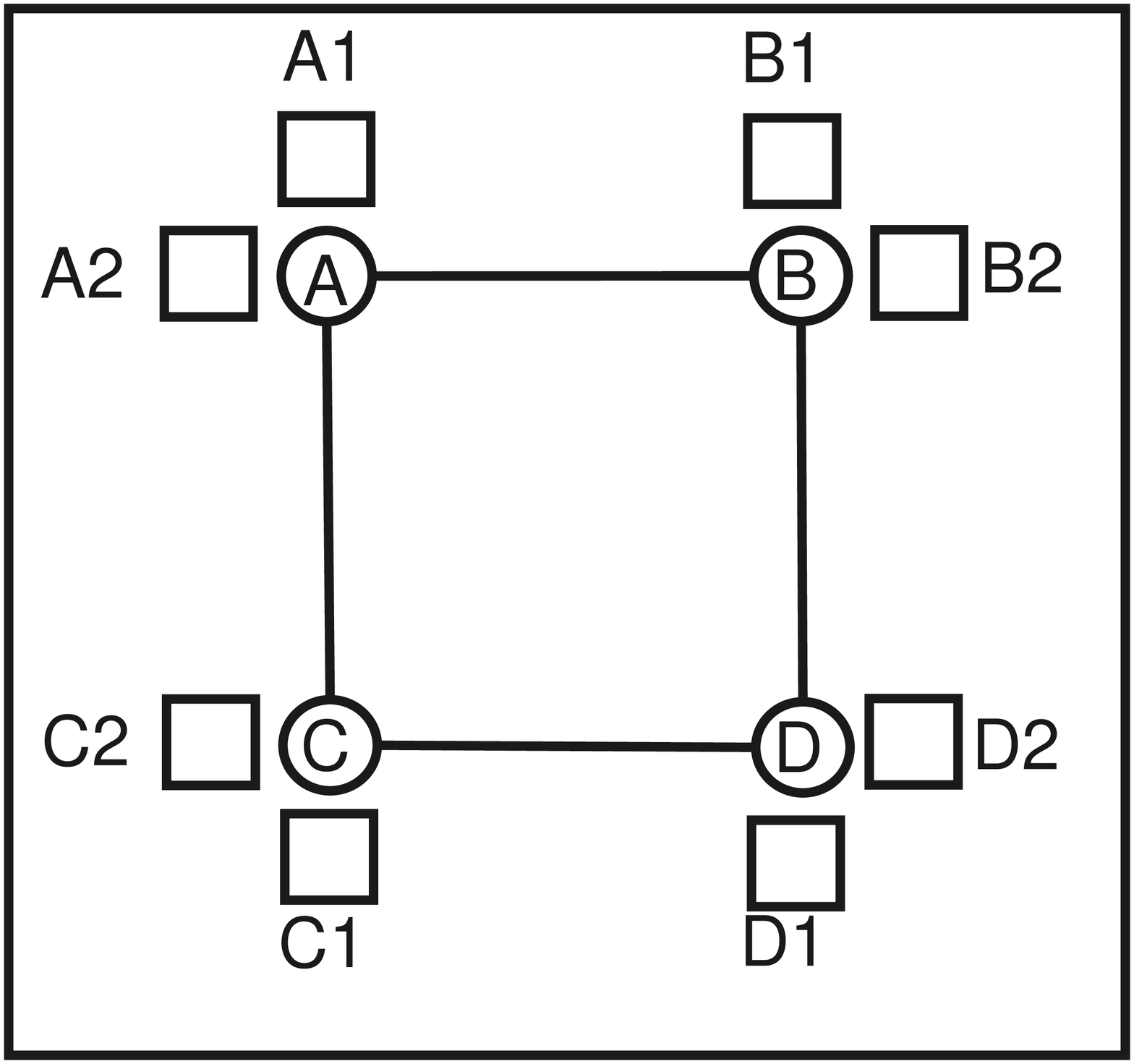}}  \hfill%
   \subfloat[Grid status after Step 1]{\includegraphics[width=.4\linewidth]{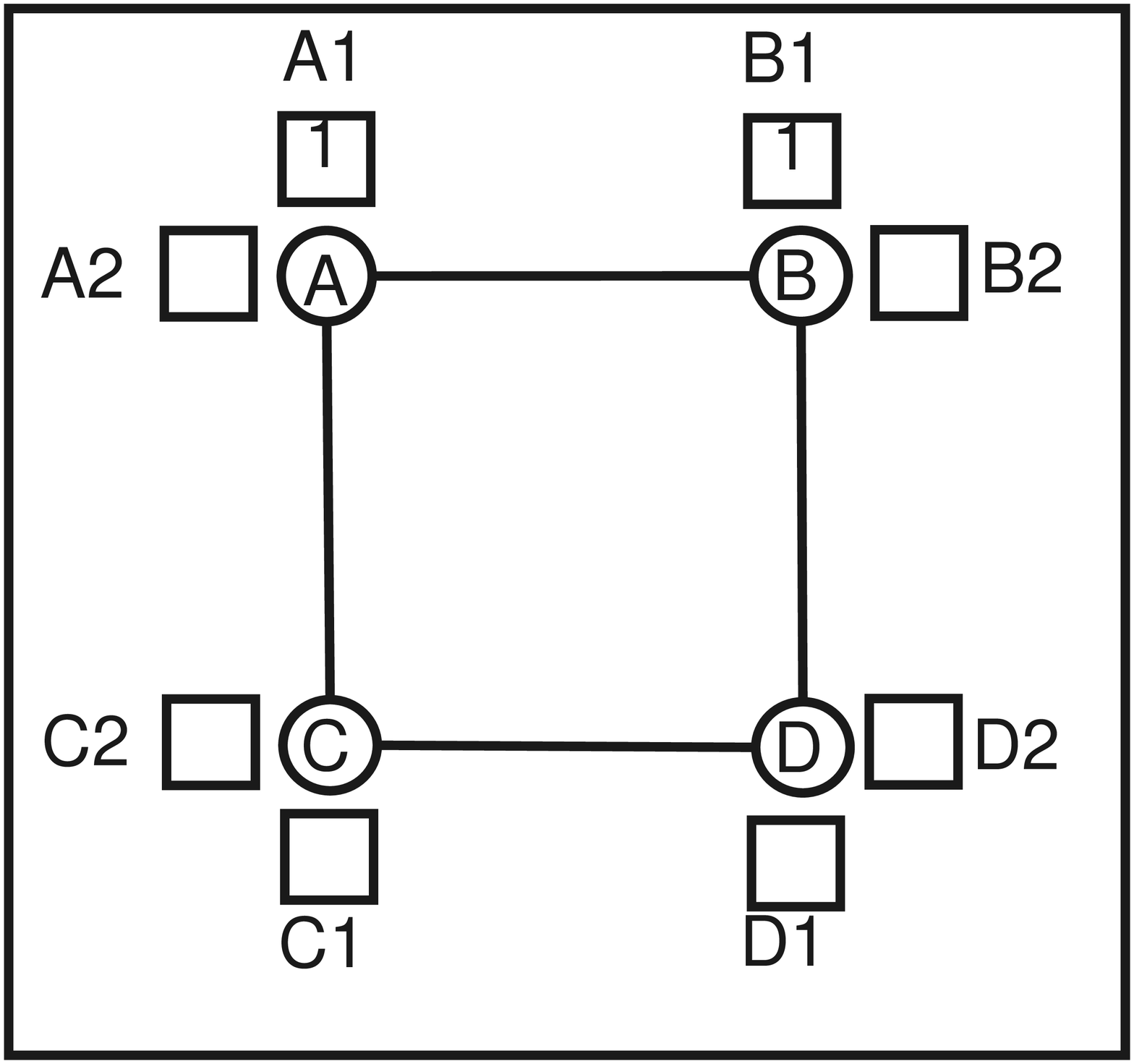}}%
    \end{tabular}
    \caption{NOCAG at initial stage and step 1 }
     \label{toy_2}
\end{figure}

Consider the network topology shown in the Figure~\ref{toy_2}~(a). It has four nodes A, B, C and D with each node having a pair of radios for communication. 
Let $A_1$, $A_2$ be the radios on node A similarly $B_1$, $B_2$ on B , $C_1$,$C_2$ on C and $D_1$, $D_2$ on D.
Let's say we have three available channels $Ch_1$,$Ch_2$ and $Ch_3$. 

\begin{figure}
  \centering%
  \begin{tabular}{cc}
  \subfloat[Grid status after Step 2]{\includegraphics[width=.4\linewidth]{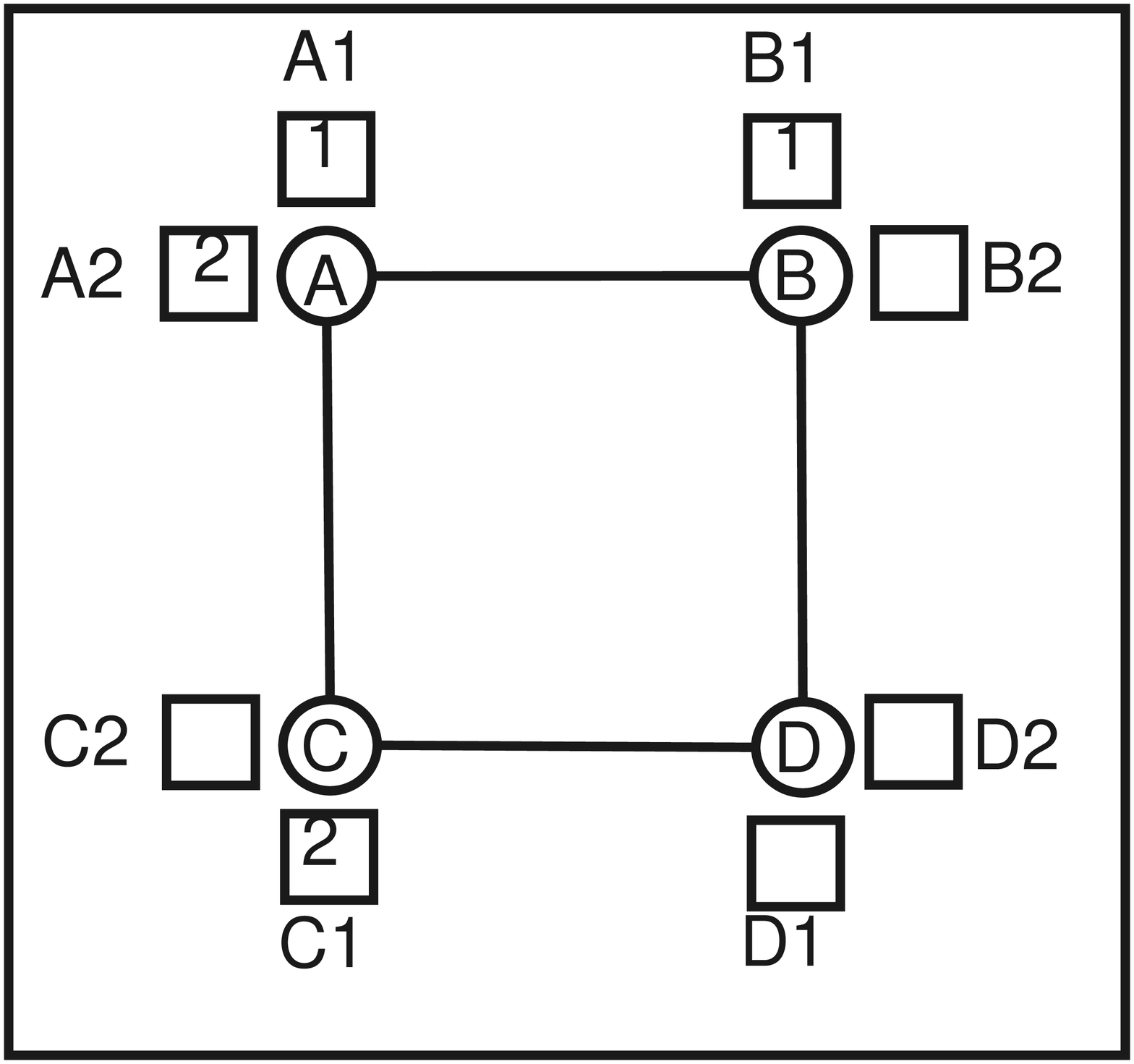}}  \hfill%
   \subfloat[Grid status after Step 3]{\includegraphics[width=.4\linewidth]{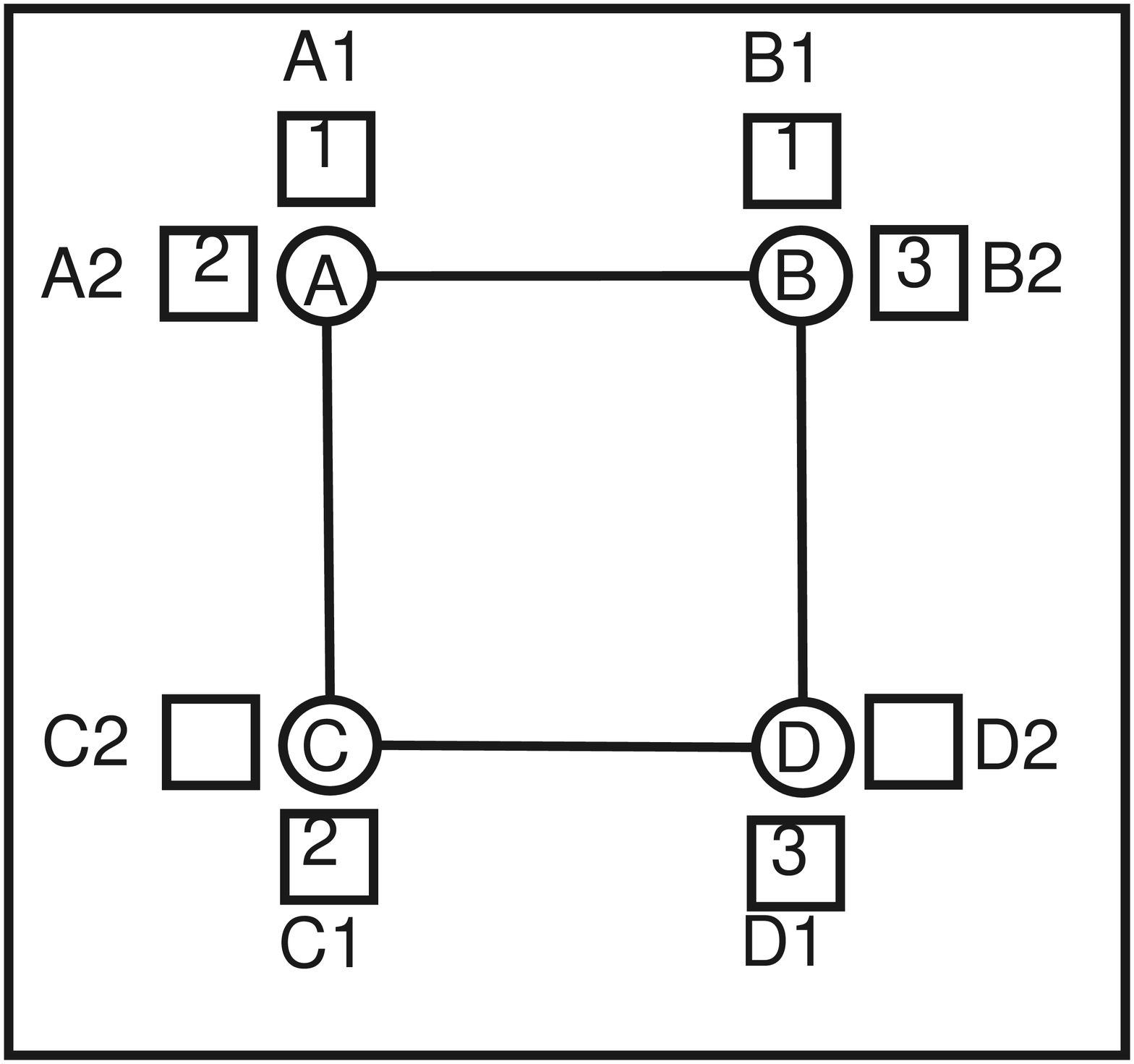}}%
    \end{tabular}
    \caption{NOCAG at steps 2 and 3}
     \label{toy_3}
\end{figure}

\begin{figure}
  \centering%
  \begin{tabular}{cc}
  \subfloat[Grid status after Step 4]{\includegraphics[width=.4\linewidth]{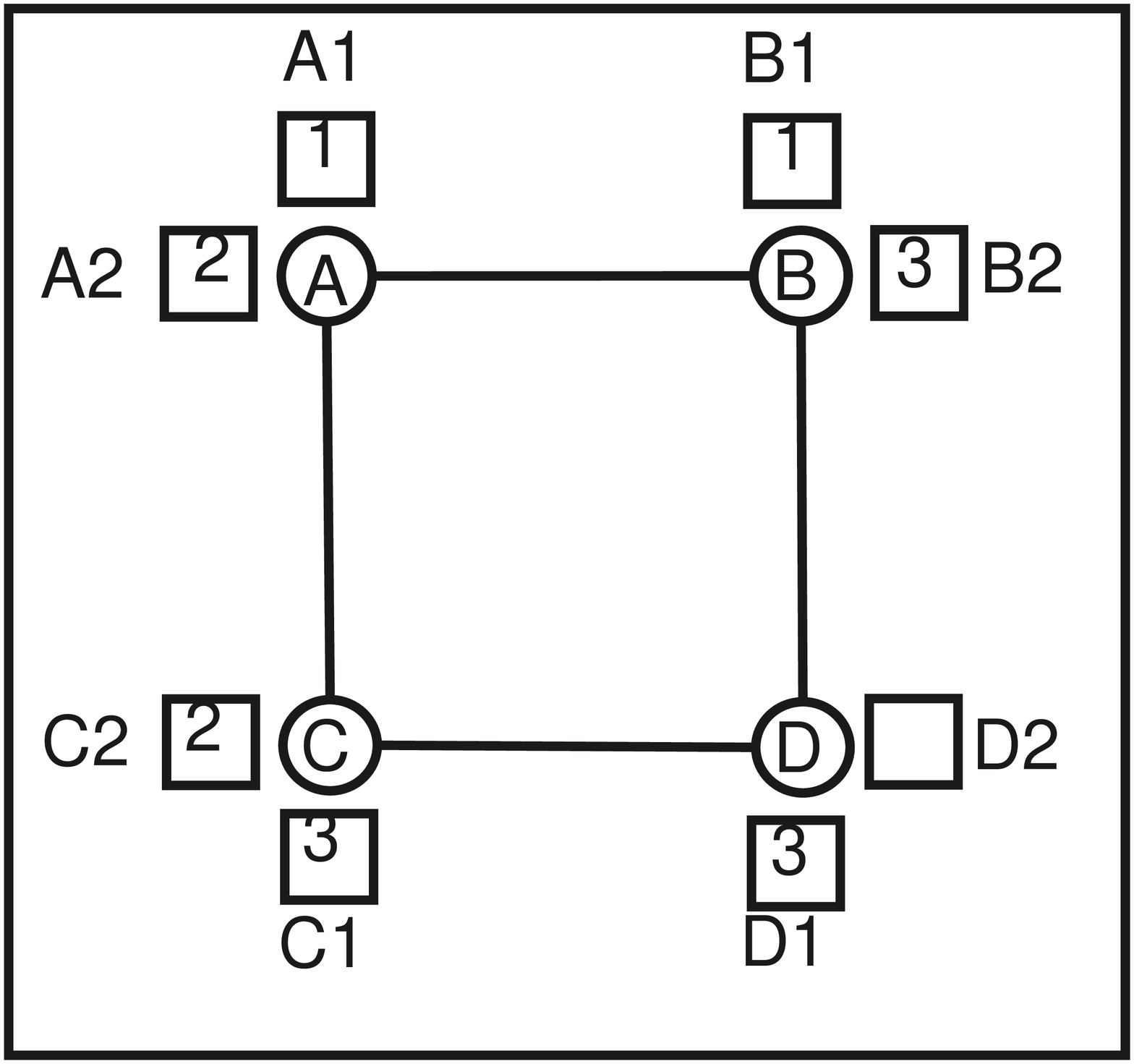}}  \hfill%
   \subfloat[Grid status after Step 5]{\includegraphics[width=.4\linewidth]{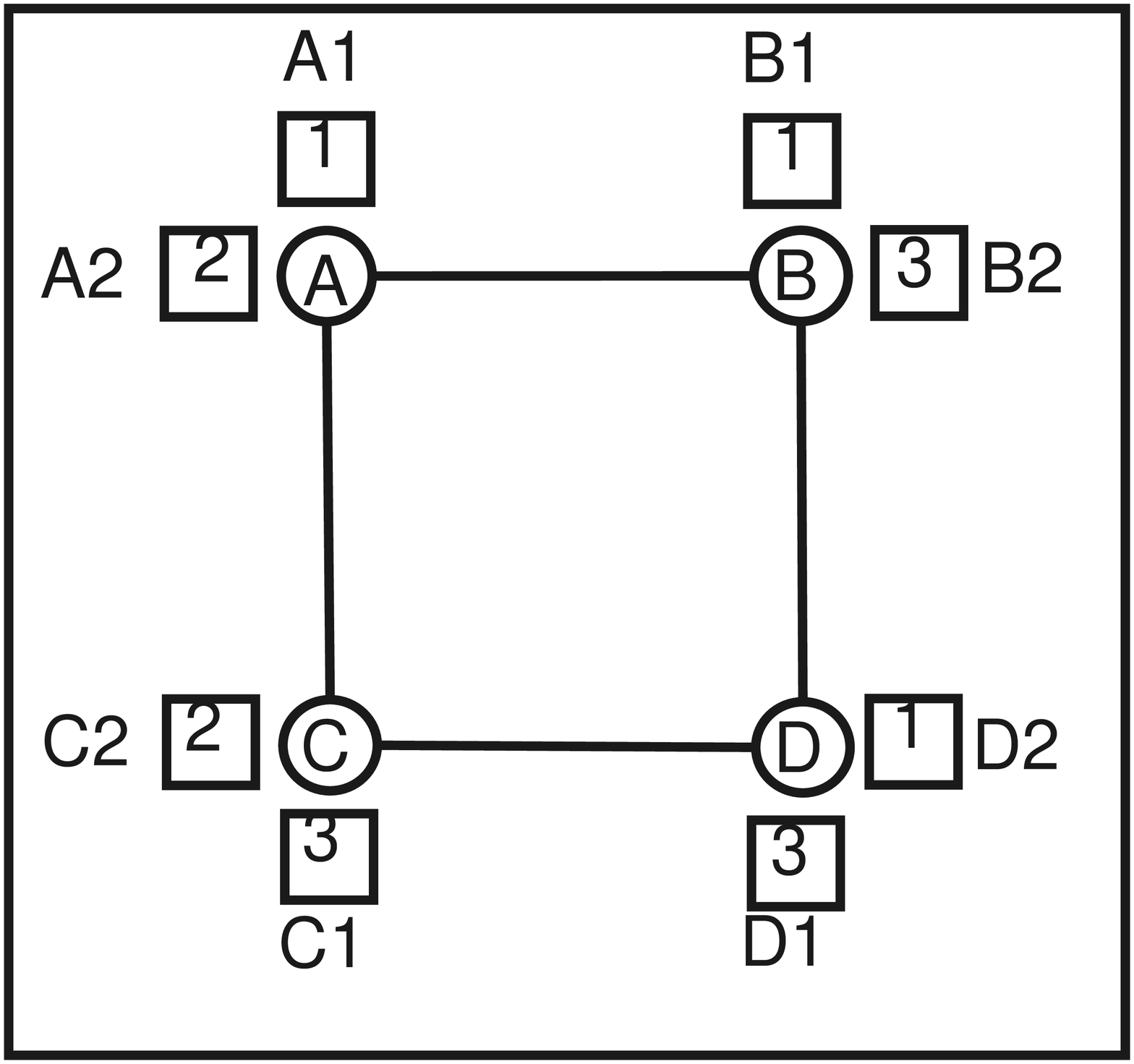}}%
    \end{tabular}
    \caption{NOCAG at steps 4 and 5}
     \label{toy_4}
\end{figure}

For a better understanding of the algorithm we have shown how the algorithm progresses in Figures~\ref{toy_2},~\ref{toy_3},~\ref{toy_4}.
A node is depicted as a circle and its name is written inside the circle. 
Boxes just besides the circle are the radios on the node named just beside the boxes.
Numbers inside the box are the channels assigned to the radios. 
For clarity purpose $Ch_1$ is depicted as 1 in the figures.

The initial state of the WMN would be look like Figure~\ref{toy_2}~(a).
Next it takes all the connected pairs of nodes and assigns channel based on the procedure described in previous section.

Let's suppose it has chooses node A and B is the node adjacent to A, so nodes A and B both have an unassigned radio and both do not have a common channel so it assigns a common channel to one of radios on each node. 
Lets suppose it has assigned channel $Ch_1$ to the radios $A_1$ and $B_1$.
After the first pair is assigned channel the status of various nodes and radios on them are 
$A_1$ $\leftarrow$ $Ch_1$, $B_1$  $\leftarrow$ $Ch_1$.
$A_2$, $B_2$,  $C_1$, $C_2$,  $D_1$, $D_2$ are still unassigned as shown in Figure~\ref{toy_2}~(b).

Now the node A has C as another adjacent node. 
So, nodes A and C both have an unassigned radio and both don't have a common channel. 
So it assigns channel $Ch_2$ for the radios on these nodes.
$A_1$ $\leftarrow$ $Ch_1$, $B_1$  $\leftarrow$ $Ch_1$,
$A_2$ $\leftarrow$ $Ch_2$, $C_1$  $\leftarrow$ $Ch_2$.
$B_2$, $C_2$, $D_1$, $D_2$ are still unassigned as shown in Figure~\ref{toy_3}~(a).

Let the next node considered is B and since A has already been considered D is the only adjacent node left. 
Nodes B and D both have an unassigned radio as earlier so the assignment shall be in the same manner.
So it chooses $Ch_3$.

So after the assignment the status of various nodes and radios, assignment would look like 

$A_1$ $\leftarrow$ $Ch_1$, $B_1$  $\leftarrow$ $Ch_1$, 
$A_2$ $\leftarrow$ $Ch_2$, $C_1$  $\leftarrow$ $Ch_2$,
$B_2$ $\leftarrow$ $Ch_3$, $D_1$  $\leftarrow$ $Ch_3$.
$C_2$, $D_2$ are still unassigned, CA after this would look like Figure~\ref{toy_3}~(b).

Now let's consider C  and since A is already considered so, D is only left adjacent node.
Both have an unassigned radio but this the $C_2$ radios must be assigned $Ch_3$ because other two cannot be assigned. 
So it assigns $Ch_3$ to $C_2$, now D has a radio that is assigned to $Ch_3$ so nothing has to changed on D. So after this CA is
$A_1$ $\leftarrow$ $Ch_1$, $B_1$  $\leftarrow$ $Ch_1$, 
$A_2$ $\leftarrow$ $Ch_2$, $C_1$  $\leftarrow$ $Ch_2$, 
$B_2$ $\leftarrow$ $Ch_3$, $D_1$  $\leftarrow$ $Ch_3$, 
$C_2$ $\leftarrow$ $Ch_3$. $D_2$ is still unassigned, CA after this would look like Figure~\ref{toy_4}~(a). 

Now it assigns channel to the only left unassigned radio a channel that least increases the interference. This is an optional step to follow at the end. So the final CA would be

$A_1$ $\leftarrow$ $Ch_1$, $B_1$  $\leftarrow$ $Ch_1$, 
$A_2$ $\leftarrow$ $Ch_2$, $C_1$  $\leftarrow$ $Ch_2$, 
$B_2$ $\leftarrow$ $Ch_3$, $D_1$  $\leftarrow$ $Ch_3$, 
$C_2$ $\leftarrow$ $Ch_3$, $D_2$  $\leftarrow$ $Ch_1$, shown in Figure~\ref{toy_4}~(b).

\subsection{Mathematical Formulation}
Having discussed how the algorithm works, we now propose a mathematical model to find the maximum achievable throughput.
We formulate the capacity problem in the  network as a Mixed-Integer Linear Programming (MILP) problem.

To formulate the problem using MILP, we need to know  about the variables used in the formulation, and constraints that bind them, to have an objective.
We now enumerate the variables followed by the constraints, and then proceed with the objective to formulate the problem.

\subsubsection{Variables}
  \begin{itemize}
      \item $flow(i,j)$ - variable denoting the amount of flow, flowing from node $i$ to node $j$, on link connecting $i$ and $j$.
      \item $C(i,j)$ - the maximum rate at which the link between node $i$ and node $j$ can transfer the data.
      \item $Rad_{max}$ - number of maximum radios on any node.
      \item $int$ - represents an intermediate node in a path from source to sink.
      
 \end{itemize}

\subsubsection{Constraints}
\begin{itemize}
 \item $Continuity$
 
The sum of all the incoming flows must be equal to the sum of all the outgoing flows at each intermediate node.
Consider the node A in the Figure~\ref{sample}, the incoming flows are $flow1$, $flow2$ and the outgoing flows are $flow3$, $flow4$.
So,  $flow1$ + $flow2$ = $flow3$ + $flow4$.

\item $Flow$

The flow on any link is non negative.
\begin{center}
$flow(i,j)$ $\geq$ $0$ 
 
\end{center}

\item $Link$ $Capacity$ 

We consider a theoretical capacity of 54 Mbps for a 802.11g link. But the actual maximal achievable capacity on any link is 9.1 Mbps \cite{throughputmax} 
as we are taking RTS/CTS into consideration, and a full TCP ACK requires more time . 
\begin{center}
 $C(i,j)$  $\leq$ $9.1 Mbps$ 
\end{center}

\item $Data$ $Validation$

Total data sent by the source nodes must be greater than or equal to the data received at sinks. We consider the greater than constraint to account for lost packets.
\begin{center}

$\sum_{i}$ $Source_i$ $\geq$ $\sum_{j}$ $Sink_j$
 
\end{center}
\item $Number$ $of$ $channels$

It is a fair assumption that the number of channels are more than number of radios taking RCI into consideration.
\begin{center}
 $Number$ $of$ $channels$ $\geq$ $Rad_{max}$
\end{center}

\end{itemize}

\subsubsection{Objective}
The objective is to maximize the throughput in the network.
\begin{equation}
\begin{split}
Maximize \sum_{k} y_{k} 
  \end{split}
\end{equation}
Where $y_{k}$ is the throughput of flow between a source-sink pair. And  $k$ denotes the source-sink pairs in the network.
\begin{equation}
\begin{split}
y_{k} = { \frac{1}{|P^k|} \sum_{i} P_i^k}  
\end{split}
\end{equation}
$P_i^k$ denotes $i^{th}$ possible path between source-link pair $k$.
An example $P_i^k$ is shown in Figure~\ref{sample} with a common source, two different sinks and a possible path through intermediate nodes shown for each source-link pair.
So to maximize $y_k$ we have to maximize flow in $P_i^k$.

\begin{equation}
\begin{aligned}
 \max P_i^k &= \min \{flow_{max}(source,int_1),  \dots ,\\
	     &\hspace{1cm} flow_{max}(int_n,sink)\}  \\
	    &= \min \{C(source,int_1),  \dots ,C(int_n,sink)\} 
\end{aligned}
 \end{equation}

This follows from our assumption that
the weakest possible link in the path is transmitting at max data rate possible \cite{ahlswede2000network}.
\begin{equation}
 \begin{split}
\max \sum_{k} y_k = \sum_{k} \frac{1}{|P_i^k|} \Big( \sum_{i} \max P_i^k   \Big)
 \end{split}
\end{equation}

\section{Simulations, Results and Analysis}
It is important to show the relevance of the NOCAG proposed with a valid experimental setup. 
We have conducted extensive simulations to study the performance of the algorithm.

To compare the performance of the algorithm we have chosen Elevated Interference Zone Mitigation (EIZM) \cite{kala2015radio} 
and we have computed the BFCA for smaller grids which tend to be best CA  for any WMN. 
EIZM is one of the recent algorithms proposed for CA to WMNs and BFCA can only be computed for smaller grids because of the high computational cost. 

And we chosen CCA described in \cite{raniwala2004centralized} as it is a grid based CA algorithm which uses the concept of conflict graph and 
TID \cite{22Ramachandran} to evaluate interference and to assign channels. 
It first sorts the links that contributes most to interference and assigns channels to the radios of the links in that order.

\subsection{Theoretical Performance Analysis}
We now analyze the performance of the algorithm for grids of various sizes choosing $CXLS_{wt}$ as the metric to evaluate interference.
$CXLS_{wt}$ is the most reliable metric proposed in \cite{kala2015reliable} for evaluating the performance of a CA. 
More the $CXLS_{wt}$ of a CA, the better is the performance of the CA.
The results are depicted in Table~\ref{CXLS}.

 \begin{table} [ht!]
   \centering

\raggedright
 \begin{tabularx}{0.5\textwidth}{|*{5}{>{\centering\arraybackslash}X|}} 
 \hline
 \backslashbox{ Grid \kern-2em}{\kern-2em CA}  & NOCAG & BF & EIZM  & CCA            \\ [0.5ex] 
 \hline\hline
 3x3                     & 14 & 15 & 11       &     8.5   \\ 
 \hline
 4x4                     & 34 & 36 & 28.5      &   15        \\
 \hline
 5x5                     & 62 & 68 & 50.5      &  33      \\
 \hline
 6x6                     & 98 & 107 & 67.5    &  60.5      \\
 \hline
 7x7                     & 142 & 151 & 96      &83          \\ [1ex] 
 \hline
\end{tabularx}

\caption{Comparison of CAs with $CXLS_{wt}$ metric}
\label{CXLS}
\end{table}
We can clearly see that NOCAG outperforms EIZM, CCA over all the grids and performs very closer to the BFCA. 

\subsection{Channel Fairness Analysis}
It is always a good idea to use all the available channels evenly \cite{kala2016interference}. 
That has been the worse problem in case of the CCA. 
We have presented statistical evenness of the CAs in Table~\ref{even}.
$x:y:z$ in the Table represents channels $Ch_1$, $Ch_2$, and $Ch_3$ are used $x$,$y$, and $z$ times respectively.
 \begin{table} [h!]

\raggedright
   \center 
 \begin{tabularx}{0.5\textwidth} {|*{5}{>{\centering\arraybackslash}X|}} 
 \hline
 \backslashbox{ Grid \kern-2em}{\kern-2em CA}  & NOCAG & BF & EIZM  & CCA            \\ 
 \hline\hline
 3x3                     & 06:06:06 & 06:06:06 & 07:05:06      &     08:01:09   \\ 
 \hline
 4x4                     & 09:11:12 & 11:11:10 & 11:09:12      &   16:08:08        \\
 \hline
 5x5                     & 15:17:18 & 16:17:17 & 16:15:19      &  25:07:18      \\
 \hline
 6x6                     & 22:24:26 & 24:24:24 & 21:21:30    &  36:12:24      \\
 \hline
 7x7                     & 31:33:35 & 33:33:32 & 32:28:38     &47:13:38          \\ 
 \hline
\end{tabularx}

\caption{ Channel Fairness}
\label{even}
\end{table}
BFCA has the best statistical evenness in channels i.e., it uses all the available channels evenly.
We can observe that NOCAG is very efficient and EIZM is also closer in statistical evenness but we can observe that CCA is not that good in
 choosing the channels efficiently thereby leading to under usage of certain channels and over usage of other channels.

\subsection{Test Scenario Developed}
We have developed a test scenario that includes each and every node for data transmission in the WMN. 
Consider a $n \times n$ grid we have $2n$ concurrent flows, $n$ vertical flows one each from top node to the bottom and $n$ horizontal flows one each directing from left most node to the rightmost node. 
This way the nodes are exhaustively used to assess the performance of the channel assignment. 

Each flow transmits a data file from the source to the sink. 
For example, top left node transmits data in a file to the bottom left node as per the vertical flow is concerned and sends data to the top right nodes as the horizontal flow in concerned. 
An example horizontal flow and vertical flow are shown in Figure~\ref{sample}.

To test the performance of the WMN we have considered Overall Throughput of the network along with Mean Delay (MD) and Packet Loss Ratio (PLR) as metrics.
So we have developed two sets of simulating environments using Transmission Control Protocol (TCP) and User Datagram Protocol (UDP) as the transport layer protocols. 
We have used the NS-3 inbuilt BulkSendApplicaton for TCP and UdpClientServer application for UDP. 
TCP simulations are aimed at finding the Network Throughput (Aggregate of all the individual flows) and the UDP simulations are aimed at finding the PLR and MD.

 \subsection{Simulation Parameters}

We have developed experimental setup in NS-3 \cite{henderson2008network} to test the performance of the algorithm on various grids of sizes varying from 3$\times$3 to 7$\times$7,
and ran extensive simulations to test the performance practically.

A data file of 5MB is sent from source to sink with parameters being 2 radios per node with a node separation of 250 mts and 3 available orthogonal channels at 2.4 GHz with IEEE protocol standard
to be 802.11g with RTS/CTS enabled.

For the UdpClientServer application, we have considered the packet size to be 1KB with a packet interval of 50ms and for the TCP BulkSendApplicaton, the maximum segment size is 1KB.
The routing protocol used is OLSR and constant as the rate control algorithm. For 802.11g maximum PHY. Datarate is 54 Mbps with MAC Fragmentation threshold to be 2200Bytes.

\subsection{GAMS Solver}
We use the General Algebraic Modeling System (GAMS) \cite{gams} Solver to model the mathematical formulation described in proposed work section.
To find the maximum achievable throughput we have assumed that every link is fully utilized without any impact of interference.
With the above described constraints, we have developed GAMS Solver and the results are presented in the Table~\ref{GAMS}.

\subsection{Experimental Results}
Thorough simulations are being run in NS-3 for the above mentioned test cases and the results are presented for exhaustive analysis of the proposed algorithm. 

\subsubsection{Throughput Analysis}
\begin{figure}
  \centering%
  {\includegraphics[width=0.65\linewidth]{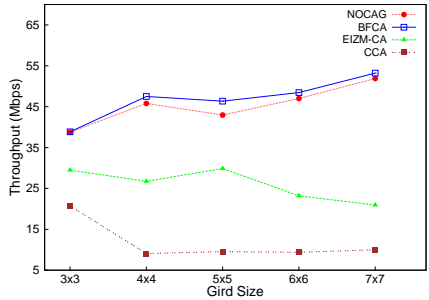}}
    \caption{Throughput Analysis }
     \label{tput}
\end{figure}

Aggregate Throughput (in Mbps) results are presented in Figure~\ref{tput} and in Table~\ref{GAMS}. We can see that NOCAG clearly outperforms EIZM and CCA 
and it is clear that NOCAG performs closer to BFCA. 
NOCAG is only 7.3\% worse than BFCA in the case of 5$\times$5 grid, and is considerably 43.8\% better than EIZM, and an overwhelming 350\% better than CCA
and similar results are observed in all the other grids.

We observe that the optimal performance achieved through BFCA is validated by the theoretical network model. 
Thus BFCA can be used as the optimal reference CA against which we compare performance of NOCAG.

\begin{table} [h!]
\tabcolsep=0.11cm
\begin{tabularx}{0.5\textwidth}{|*{4}{>{\centering\arraybackslash}X|}} 
\hline 
Grid Size& MILP Maximum Value & BF Experimental Value  & NOCAG Experimental Value\\
\hline
\hline  
3$\times$3&  		54.6		& 38.87		&38.74\\
\hline  
4$\times$4&		72.8		&47.50 		&45.80\\
\hline  
5$\times$5&		91		&46.36		&42.97\\
\hline
6$\times$6&		109.2		&48.46		&47.00\\
\hline
7$\times$7&		127.4		&53.21		&51.90\\
\hline  
\end{tabularx} 

\caption{Grid throughput of MILP, BFCA and NOCAG in Mbps}
\label{GAMS}
\end{table}
\subsubsection{Mean Delay (MD) Analysis}
\begin{figure}
  \centering%
  {\includegraphics[width=0.65\linewidth]{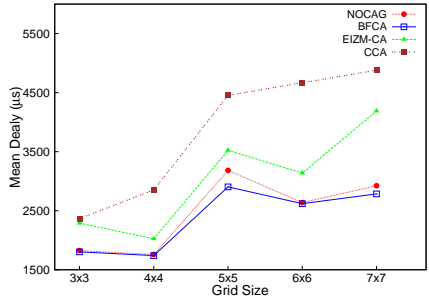}}
    \caption{MD Analysis }
     \label{md}
\end{figure}

Mean Delay (in microseconds) results are shown in Figure~\ref{md}. 
We can observe that NOCAG results are better than EIZM and CCA in Mean Delay as the metric as well. 
BFCA and NOCAG perform closer as we can see the results.
NOCAG is only 9.6\% worse than BFCA in case of 5$\times$5 grid, and is considerably 9.6\% better than EIZM, and  28\% better than CCA
and similar results are observed in all the grids.

\subsubsection{Packet Loss Ratio (PLR) Analysis}

\begin{figure}
  \centering%
  {\includegraphics[width=0.65\linewidth]{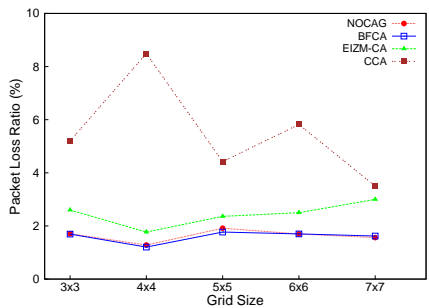}}
    \caption{PLR Analysis}
     \label{plr}
\end{figure}

We can observe the PLR  results in Figure~\ref{plr}. 
Considering PLR as the metric also, NOCAG clearly is performing better than EIZM and CCA. 
Notable point is that NOCAG performs near to  BFCA. 
NOCAG is only 8.1\% worse than BFCA in the case of 5$\times$5 grid, and  is considerably 19\% better than EIZM, and  34\% better than CCA. 
And similar patterns are observed in all the grids.

CCA has a real low performance because of the reason of uneven channel fairness, and it does not take RCI into consideration. 

\section{conclusions}
Having discussed in detail how the algorithm works and its performance, we can now make some valuable conclusions. 
It is clear that the performance of a WMN changes drastically with the CA. 
The performance of a WMN includes the network throughput, PLR, and MD though are not the only exhaustive metrics but these deserve evaluation.
It is clear from the plots that a good CA would effect all the performance metrics.

It is imperative to say that the NOCAG has a very low computational overhead of just linear in terms of  number of nodes O(m), where m is the number of nodes in the WMN when compared to 
the exponential high computational overhead of BFCA.
Having seen the results, it makes sense that we proposed a fairly better channel assignment algorithm that
outperforms previous best algorithms, and performs as good as to BFCA.

\bibliography{Ok}
\end{document}